\documentclass[conference,12pt]{IEEEtran}
\usepackage{epsfig}
\usepackage{graphicx}
\usepackage{cite}
\usepackage{amsmath}
\usepackage{amssymb}

\newcommand{\ignore}[1]{}
\newcommand{\comment}[1]  {}
\DeclareMathOperator*{\argmax}{arg\,max}
\newcommand\ie{{\textsl{i.e.\,}}}

\newcommand\etal{{\textsl{et al.\,}}}
\def\BE{\begin{equation}}
\def\EE{\end{equation}}
\def\BEA{\begin{eqnarray}}
\def\EEA{\end{eqnarray}}
\newcommand{\cut}[1]{{}}
\newcommand\vb{{\bf b}}
\newcommand\vc{{\bf c}}

\newcommand\ve{{\bf e}}

\newcommand\vr{{\bf r}}

\newcommand\vw{{\bf w}}
\newcommand\vx{{\bf x}}
\newcommand\vy{{\bf y}}

\newcommand\mA{{\bf A}} 

\newtheorem{thm}{Theorem}

\begin{document}
\twocolumn
%
\title{Distributed Kalman Filter \\via Gaussian Belief Propagation}

\author{\IEEEauthorblockN{Danny Bickson}
\IEEEauthorblockA{IBM Haifa Research Lab\\
Mount Carmel, Haifa 31905, Israel\\
Email: dannybi@il.ibm.com}
 \and \IEEEauthorblockN{Ori Shental
} \IEEEauthorblockA{Center for Magnetic\\ Recording
Research\\UCSD\\9500 Gilman Drive\\La Jolla, CA 92093, USA\\Email:
oshental@ucsd.edu} \and \IEEEauthorblockN{Danny Dolev}
\IEEEauthorblockA{School of Computer Science\\and
Engineering\\Hebrew University of Jerusalem\\Jerusalem 91904,
Israel\\Email: dolev@cs.huji.ac.il}}

\maketitle

\begin{abstract}
Recent result shows how to compute distributively and efficiently
the linear MMSE for the multiuser detection problem, using the
Gaussian BP algorithm. In the current work, we extend this
construction, and show that operating this algorithm twice on the
matching inputs, has several interesting interpretations. First,
we show equivalence to computing one iteration of the Kalman
filter. Second, we show that the Kalman filter is a special case
of the Gaussian information bottleneck algorithm, when the weight
parameter $\beta = 1$. Third, we discuss the relation to the
Affine-scaling interior-point method and show it is a special case
of Kalman filter.

Besides of the theoretical interest of this linking estimation,
compression/clustering and optimization, we allow a single
distributed implementation of those algorithms, which is a highly
practical and important task in sensor and mobile ad-hoc networks.
Application to numerous problem domains includes collaborative
signal processing and distributed allocation of resources in a
communication network.

\end{abstract}

\section{Introduction}
Recent work \cite{ISIT2} shows how to compute efficiently and
distributively the MMSE prediction for the multiuser detection
problem, using the Gaussian Belief Propagation (GaBP) algorithm.
The basic idea is to shift the problem from linear algebra domain
into a probabilistic graphical model, solving an equivalent
inference problem using the efficient belief propagation inference
engine. \cite{Allerton} compares the empirical performance of the
GaBP algorithm relative to other linear iterative algorithms,
demonstrating faster convergence. \cite{ISIT1} elaborates on the
relation to solving systems of linear equations.

In the present work, we propose to extend the previous
construction, and show, that by performing the MMSE computation
twice on the matching inputs we are able to compute several
algorithms. First, we reduce the discrete Kalman filter
computation \cite{Kalman} to a matrix inversion problem and show
how to solve it using the GaBP algorithm. We show that Kalman
filter iteration which is composed from prediction and measurement
steps can be computed by two consecutive MMSE predictions. Second,
we explore the relation to Gaussian information bottleneck (GIB)
\cite{GIB} and show that Kalman filter is a special instance of
the GIB algorithm, when the weight parameter $\beta = 1$. To the
best of the authors knowledge, this is the first algorithmic link
between the information bottleneck framework and linear dynamical
systems. Third, we discuss the connection to the Affine-scaling
interior-point method and show it is an instance of the Kalman
filter.

Besides of the theoretical interest of linking compression,
estimation and optimization together, our work is highly practical
since it proposes a general framework for computing all of the
above tasks distributively in a computer network. This result can have many
applications in the fields of estimation, collaborative signal
processing, distributed resource allocation etc.

A closely related work is \cite{BibDB:FactorGraph}.
In this work, Frey \etal focus on the belief propagation algorithm
(a.k.a sum-product algorithm) using factor graph topologies. They
show that the Kalman filter algorithm can be computed using belief
propagation over a factor graph. In this contribution we extend
their work in several directions. First, we extend the computation
to vector variables (relative to scalar variables in Frey's work).
Second, we use a different graphical model: an undirected
graphical model which results in simpler update rules, where Frey
uses factor-graph with two types of messages: factor to variables
and variables to factors. Third and most important, we allow an
efficient distributed calculation of the Kalman filter steps,
where Frey's algorithm is centralized.

Another related work is \cite{KalmanKarmarkar}. In this work the
link between Kalman filter and linear programming is established.
In this work we propose a new and different construction which
ties the two algorithms.

The structure of this paper is as follows. In Section \ref{sec_kf}
we describe the discrete Kalman filter. In Section \ref{sec_GIB}
we outline the GIB algorithm and discuss its relation to the
Kalman filter. Section \ref{sec_as} presents the Affine-scaling
interior-point method and compares it to the Kalman filter
algorithm. Section \ref{sec_GaBP} presents our novel construction
for performing an efficient distributed computation of the three
methods.

\section{Kalman Filter}
\label{sec_kf}
\subsection{An Overview}
 The Kalman filter is an efficient iterative
algorithm to estimate the state of a discrete-time controlled
process $x \in R^n$ that is governed by the linear stochastic
difference equation \footnote{In this paper, we assume there is no
external input, namely $x_k = Ax_{k-1} + w_{k-1}$. However, our
approach can be easily extended to support external inputs.}: \BE
x_k = Ax_{k-1} + Bu_{k-1} + w_{k-1}, \EE \

with a measurement $z \in R^m$ that is $ z_k = Hx_k + v_k.$ The
random variables $w_k$ and $v_k$ that represent the process and
measurement AWGN noise (respectively). $p(w) \sim \mathcal{N}(0,
Q), p(v) \sim \mathcal{N}(0, R)$. We further assume that the
matrices $A,H,B,Q,R$ are given\footnote{Another possible extension
is that the matrices $A,H,B,Q,R$ change in time, in this paper we
assume they are fixed. However, our approach can be generalized to
this case as well.}.

The discrete Kalman filter update equations are given
by~\cite{Kalman}:

The prediction step:
\begin{subequations}
\begin{eqnarray}
 \hat{x}^-_k &=& A\hat{x}_{k-1} + Bu_{k-1}, \label{hat_x_minus_k}\\
 P_k^- &=& AP_{k-1}A^T + Q. \label{p_k_minus}
\end{eqnarray}
\end{subequations}

The measurement step:
\begin{subequations}
\begin{eqnarray}
 K_k &=& P_k^-H^T(HP_k^-H^T + R)^{-1}, \label{kalman_gain} \\
\hat{x}_k &=& \hat{x}^-_k + K_k(z_k - H\hat{x}^-_k),
\label{hat_x_k} \\
 P_k &=& (I-K_kH)P_k^-. \label{p_k}
\end{eqnarray}
\end{subequations}
where $I$ is the identity matrix.

The algorithm operates in rounds. In round $k$ the estimates
$K_k,\hat{x}_k,P_k$ are computed, incorporating the (noisy)
measurement $z_k$ obtained in this round. The output of the
algorithm are the mean vector $\hat{x}_k$ and the covariance
matrix $P_k$.

\subsection{New construction} \label{new-cons} Our novel
contribution is a new efficient distributed algorithm for
computing the Kalman filter. We begin by showing that the Kalman
filter can be computed by inverting the following covariance
matrix: \BE E =
\left(%
\begin{array}{ccc}
  -P_{k-1} & A & 0 \\
  A^T & Q & H \\
  0 & H^T & R \\
\end{array}%
\right)\,, \label{mat_E} \EE and taking the lower right $1 \times 1$
block to be $P_k$.

The computation of $E^{-1}$ can be done efficiently using recent
advances in the field of Gaussian belief propagation
\cite{ISIT1,ISIT2}. The intuition for our approach, is that the
Kalman filter is composed of two steps. In the prediction step,
given $x_k$, we compute the MMSE prediction of $x_k^-$
\cite{BibDB:FactorGraph}. In the measurement step, we compute the
MMSE prediction of $x_{k+1}$ given $x_k^-$, the output of the
prediction step. Each MMSE computation can be done using the GaBP
algorithm \cite{ISIT2}. The basic idea, is that given the joint
Gaussian distribution $p(\vx,\vy)$ with the covariance matrix $
C = \left(%
\begin{array}{cc}
  \Sigma_{xx} & \Sigma_{xy} \\
  \Sigma_{yx} & \Sigma_{yy} \\
\end{array}%
\right)$, we can compute the MMSE prediction \[ \hat{y} =
\argmax_{y}p(y|x) \propto \mathcal{N}(\mu_{y|x},
\Sigma_{y|x}^{-1})\,,
\] where \[ \mu_{y|x} = (\Sigma_{yy} -
\Sigma_{yx}\Sigma_{xx}^{-1}\Sigma_{xy})^{-1} \Sigma_{yx} \Sigma_{xx}^{-1}x\,, \]
\[ \Sigma_{y|x} = (\Sigma_{yy} -
\Sigma_{yx}\Sigma_{xx}^{-1}\Sigma_{xy})^{-1}\,. \] This in turn is
equivalent to computing the Schur complement of the lower right
block of the matrix $C$. In total, computing the MMSE prediction
in Gaussian graphical model boils down to a computation of a
matrix inverse. In \cite{ISIT1} we have shown that GaBP is an
efficient iterative algorithm for solving a system of linear
equations (or equivalently computing a matrix inverse). In
\cite{ISIT2} we have shown that for the specific case of linear
detection we can compute the MMSE estimator using the GaBP
algorithm. Next, we show that performing two consecutive
computations of the MMSE are equivalent to one iteration of the
Kalman filter.
\\
\begin{thm} The lower right $1 \times 1$ block of the matrix inverse $E^{-1}$
(eq. \ref{mat_E}), computed by two MMSE iterations, is equivalent to the computation of $P_k$ done
by one iteration of the Kalman filter algorithm.\\
\end{thm}
Proof of Theorem 1 is given in Appendix A.\\
\\
In Section \ref{sec_GaBP} we explain how to utilize the above
observation to an efficient distributed iterative algorithm for
computing the Kalman filter.

\section{Gaussian Information Bottleneck}
\label{sec_GIB} Given the joint distribution of a source variable
X and another relevance variable Y, Information bottleneck (IB)
operates to compress X, while preserving information about
Y~\cite{InfoBottleneck,Slonim}, using the following variational
problem:

\[ \min_{p(t|x)} \mathcal{L} : \mathcal{L} \equiv I(X; T) - \beta I(T; Y ) \]
$T$ represents the compressed representation of $X$ via the
conditional distributions $p(t|x)$, while the information that $T$
maintains on $Y$ is captured by the distribution $p(y|t)$. $\beta
> 0 $ is a lagrange multiplier which weights the tradeoff between
minimizing the compression information and maximizing the relevant
information. As $\beta \rightarrow 0$ we are interested solely in
compression, but all relevant information about Y is lost $I(Y;T)
= 0$. When $\beta \rightarrow \infty$ where are focused on
preservation of relevant information, in this case T is simply the
distribution X and we obtain $I(T;Y) = I(X;Y)$. The interesting
cases are in between, when for finite values of $\beta$ we are
able to extract rather compressed representation of X while still
maintaining a significant fraction of the original information
about Y.

An iterative algorithm for solving the IB problem is given in
\cite{Slonim}:

\begin{subequations}
\label{IIB}
\begin{eqnarray}
P^{k+1}(t|x) =&\frac{P^k(t)}{Z^{k+1}(x,\beta)} \cdot \nonumber \\
& \cdot \exp(-\beta D_{KL}[p(y|x)||p^k(y|t)]), \label{IB_it1} \\
P^k(t)=& \int_x p(x)P^k(t|x)dx, \label{IB_it2} \\
P^k(y|t)=& \frac{1}{P^k(t)} \int_x P^k(t|x)p(x,y)dx. \label{IB_it3}
\end{eqnarray}
 \end{subequations}
where $Z^{k+1}$ is a normalization factor computed in round $k+1$.

The Gaussian information bottleneck (GIB) \cite{GIB} deals with
the special case where the underlying distributions are Gaussian.
In this case, the computed distribution $p(t)$ is Gaussian as
well, represented by a linear transformation $T_k = A_kX + \xi_k$
where $A_k$ is a joint covariance matrix of $X$ and $T$, $\xi_k \sim
\mathcal{N}(0, \Sigma_{\xi_k})$ is a multivariate Gaussian independent of X.
The outputs of the algorithm are the covariance matrices
representing the linear transformation T: $A_k, \Sigma_{\xi_k}$.

An iterative algorithm is derived by substituting Gaussian
distributions into (\ref{IIB}), resulting in the following
update rules: 
 \begin{subequations}
    \begin{eqnarray}
   \Sigma_{\xi+1}& = &(\beta\Sigma_{t_k|y} - (\beta - 1)\Sigma_{t_k}^{-1}) \label{GIB_xi}, \\
 A_{k+1} &=& \beta\Sigma_{\xi_k+1}\Sigma^{-1}_{t_k|y}A_k(I -
 \Sigma_{y|x}\Sigma_x^{-1}). \label{GIB_A}
    \end{eqnarray}
 \end{subequations}

\begin{table}[h!]
\begin{center}
 \caption{Summary of notations in the GIB
\cite{GIB} paper vs. Kalman filter \cite{Kalman}}
\begin{tabular}{|c|c|l|}
\hline \label{notations}
  GIB \cite{GIB} & Kalman \cite{Kalman} &  Kalman meaning \\
  \hline
  $\Sigma_x$ & $P_0$ & a-priori estimate error covariance \\
  $\Sigma_y$ & $Q$ & Process AWGN noise\\
  $\Sigma_{t_k}$ & $R$ & Measurement AWGN noise\\
  $\Sigma_{xy}, \Sigma_{yx}$ & $A , A^T$ & process state transformation matrix \\
  $\Sigma_{xy}A, A^T\Sigma_{yx}$ & $H^T , H$ & measurement transformation matrix \\
  $\Sigma_{\xi_k}$ & $P_k$ &  posterior error covariance in round k\\
  $\Sigma_{x|y_k}$ & $P_k^-$ & a-priori error covariance in round k\\ \hline
\end{tabular}
\end{center}
\end{table}

\begin{figure}
\begin{center}
  \includegraphics[scale=0.5, bb=2 11 390 620]{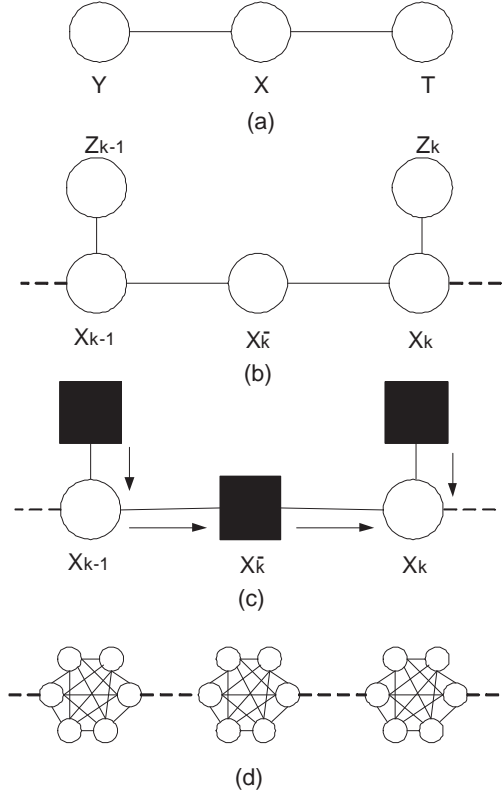}\\
  \caption{Comparison of the different graphical models used. (a) Gaussian Information Bottleneck \cite{GIB}
  (b) Kalman Filter (c) Frey's sum-product factor graph \cite{BibDB:FactorGraph} (d) Our new construction.}\label{graphical_model}
\end{center}
\end{figure}

Since the underlying graphical model of both algorithms (GIB and
Kalman filter) is Markovian with Gaussian probabilities, it is
interesting to ask what is the relation between them. In this work
we show, that the Kalman filter posterior error covariance
computation is a special case of the GIB algorithm when $\beta =
1$. Furthermore, we show how to compute GIB using the Kalman
filter when $\beta > 1$ (the case where $0 < \beta < 1$ is not
interesting since it gives a degenerate solution where $A_k \equiv
0$ \cite{GIB}.) Table \ref{notations} outlines the different
notations used by both algorithms.
\\
\\
\begin{thm}
The GIB algorithm when $\beta= 1$ is equivalent to the Kalman
filter algorithm.
\end{thm}
The proof is given in Appendix B.
\\
\begin{thm}
The GIB algorithm when $\beta > 1$ can be computed by a modified
Kalman filter iteration.
\end{thm}
The proof is given in Appendix C.\\

\begin{figure}
\begin{center}
  \includegraphics[scale=0.5]{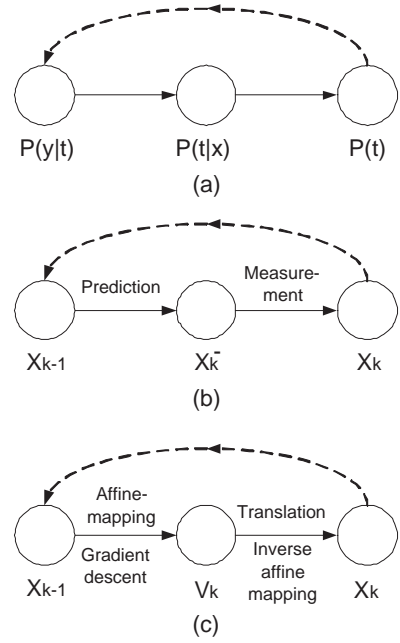}\\
  \caption{Comparison of the schematic operation of the different algorithms. (a) iterative information bottleneck operation (b) Kalman filter operation (c) Affine-scaling operation.}\label{iterative-model}
\end{center}
\end{figure}

There are some differences between the GIB algorithm and Kalman
filter computation. First, the Kalman filter has input
observations $z_k$ in each round. Note that the observations do
not affect the posterior error covariance computation $P_k$ (eq.
\ref{p_k}), but affect the posterior mean $\hat{x}_k$ (eq.
\ref{hat_x_k}).  Second, Kalman filter computes both posterior
mean $\hat{x_k}$ and error covariance $P_k$. The covariance
$\Sigma_{\xi_k}$ computed by the GIB algorithm was shown to be
identical to $P_k$ when $\beta = 1$. The GIB algorithm does not
compute the posterior mean, but computes an additional covariance
$A_k$ (eq. \ref{GIB_A}), which is assumed known in the Kalman
filter.

From the information theoretic perspective, our work extends the
ideas presented in \cite{PredictiveIB}. Predictive information is
defined to be the mutual information between the past and the
future of a time serias. In that sense, by using Theorem 2, Kalman
filter can be thought of as a prediction of the future, which from
the one hand compresses the information about past, and from the
other hand maintains information about the present.

The origins of similarity between the GIB algorithm and Kalman
filter are rooted in the IB iterative algorithm: For computing
(\ref{IB_it1}), we need to compute (\ref{IB_it2},\ref{IB_it3}) in
recursion, and vice versa.

\section{Relation to the Affine-scaling algorithm}
\label{sec_as} One of the most efficient interior point methods
used for linear programming is the Affine-scaling algorithm
\cite{Affine-scaling}. It is known that the Kalman filter is
linked to the Affine-scaling algorithm \cite{KalmanKarmarkar}. In
this work we give an alternate proof, based on different
construction, which shows that Affine-scaling is an instance of
Kalman filter, which is an instance of GIB. This link between
estimation and optimization allows for numerous applications.
Furthermore, by providing a single distribute efficient
implementation of the GIB algorithm, we are able to solve numerous
problems in communication networks.

The linear programming problem in its canonical form is given by:
 \begin{subequations}
    \begin{eqnarray}
   \mbox{minimize} & \vc^T\vx \label{can-linear} \\
    \mbox{subject to} & A\vx = \vb , \ \ \ \ \vx \ge 0. \label{can-constraints}
    \end{eqnarray}
 \end{subequations}
where $A \in \mathbb{R}^{n \times p}$ with $\mathbf{rank}\{ A \} =
p < n$. We assume the problem is solvable with an optimal $\vx^*$.
We also assume that the problem is strictly feasible, in other
words there exists $\vx \in \mathbb{R}^n$ that satisfies $A\vx =
\vb$ and $\vx > 0$.

The Affine-scaling algorithm \cite{Affine-scaling} is summarized
below. Assume $\vx_0$ is an interior feasible point to
(\ref{can-constraints}). Let $D = diag(\vx_0)$. The Affine-scaling
is an iterative algorithm which computes a new feasible point that
minimizes the cost function (\ref{can-linear}): \BE \vx_1 = \vx_0
- \frac{\alpha}{\gamma}D^2 \vr \label{af_11}  \EE where $0 <
\alpha < 1$ is the step size, $\vr$ is the step direction.
\begin{subequations}
    \begin{eqnarray}
   \vr &=& (\vc - A^T\vw), \label{af_9b} \\
   \vw &=& (AD^2A^T)^{-1}AD^2\vc, \label{af_9c} \\
   \gamma &=& \max_i(\ve_i P D \vc). \label{af_9d}
    \end{eqnarray}
 \end{subequations}
Where $\ve_i$ is the $i^{th}$ unit vector and $P$ is a projection matrix
 given by: \BE P = I - DA^T(AD^2A^T)^{-1}AD. \label{af_10} \EE
The algorithm continues in rounds and is guaranteed to find an
optimal solution in at most $n$ rounds. In a nutshell, in each
iteration, the Affine-scaling algorithm first performs an
Affine-scaling with respect to the current solution point $\vx_i$
and obtains the direction of descent by projecting the gradient of
the transformed cost function on the null space of the constraints
set. The new solution is obtained by translating the current
solution along the direction found and then mapping the result
back into the original space \cite{KalmanKarmarkar}. This has
interesting analogy for the two phases of the Kalman filter.

\begin{thm}
The Affine-scaling algorithm iteration is an instance of the
Kalman filter algorithm iteration.
\end{thm}
Proof is given in Appendix D.

\section{Efficient distributed computation}
\label{sec_GaBP} We have shown how to express the Kalman filter,
Gaussian information bottleneck and Affine-scaling algorithms as a
two step MMSE computation. Each step involves inverting a $2
\times 2$ block matrix. Recent result by Bickson and Shental \etal
\cite{ISIT2} show that the MMSE computation can be done
efficiently and distributively using the Gaussian belief
propagation algorithm. Because of space limitations the full
algorithm is not reproduced here.


The interested reader is referred to \cite{ISIT1,ISIT2} for a
complete derivation of the GaBP update rules and convergence
analysis. The GaBP algorithm is summarized in
Table~\ref{tab_summary}.

\begin{table*}[htb!]
\normalsize \caption{Computing $\vx = \mA^{-1}\vb$ via GaBP
\cite{ISIT1}.} \centerline{ \label{tab_summary}
\begin{tabular}{|c|c|l|}
  \hline
  \textbf{\#} & \textbf{Stage} & \textbf{Operation}\\
  \hline
  1. & \emph{Initialize} & Compute $P_{ii}=A_{ii}$ and $\mu_{ii}=b_{i}/A_{ii}$.\\
  && Set $P_{ki}=0$ and $\mu_{ki}=0$, $\forall k\neq i$.\\ \hline
  2. & \emph{Iterate} & Propagate $P_{ki}$ and $\mu_{ki}$, $\forall k\neq i \;
\mbox{\rm such that} \; A_{ki}\neq0$.\\& & Compute $P_{i\backslash
j}=P_{ii}+\sum_{{k}\in\mathbb{N}(i) \backslash j} P_{ki}$ and
$\mu_{i\backslash j} = P_{i\backslash
j}^{-1}(P_{ii}\mu_{ii}+\sum_{k \in \mathrm{N}(i) \backslash j}
P_{ki}\mu_{ki})$.\\
  && Compute $P_{ij} = -A_{ij}P_{i\backslash j}^{-1}A_{ji}$ and $\mu_{ij} =
-P_{ij}^{-1}A_{ij}\mu_{i\backslash j}$.\\\hline
  3. & \emph{Check} & If $P_{ij}$ and $\mu_{ij}$ did not converge, return to
    \#2. Else, continue to \#4.\\\hline
  4. & \emph{Infer} & $P_{i}=P_{ii}+\sum_{{k}\in\mathrm{N}(i)}
P_{ki}$ , $\mu_{i}=P_{i}^{-1}(P_{ii}\mu_{ii}+\sum_{k \in
\mathrm{N}(i)} P_{ki}\mu_{ki})$.\\
  \hline
  5. & \emph{Output} & $x_{i}=\mu_{i}$ \\\hline
\end{tabular}} 
\end{table*}

Regarding convergence, if it converges, GaBP is known to result in
exact inference~\cite{Weiss}. Determining the exact region of convergence and
convergence rate remain open research problems. All that is known
is a sufficient (but not necessary) condition~\cite{WS,JMLR}
stating that GaBP converges when the spectral radius satisfies
\mbox{$\rho(|I_{K}-A|)<1$}, where $A$ is first normalized s.t. the main
diagonal contains ones. A stricter sufficient condition~\cite{Weiss}, determines that
the matrix $A$ must be diagonally dominant (\ie,
$|A_{ii}|>\sum_{j\neq i}|A_{ij}| , \forall i$) in order for GaBP
to converge.

Regarding convergence speed, \cite{Allerton08-1} shows that when
converging, the algorithm converges in
$O(log(\epsilon)/log(\gamma))$ iterations, where $\epsilon$ is the
desired accuracy, and $1/2 < \gamma < 1$ is a parameter related to
the inverted matrix. The computation overhead in each iteration
is determined by the number of non-zero elements of the inverted matrix
$A$. In practice, \cite{PPNA08} demonstrates
convergence of 5-10 rounds on sparse matrices with
several millions of variables. \cite{ECCS08} shows convergence of
dense constraint matrices of size up to $150,000 \times 150,000$
in 6 rounds, where the algorithm is run in parallel using 1,024
CPUs. Empirical comparison with other iterative algorithms is given
in \cite{Allerton}.

\section{Example application}
\begin{figure}
\begin{center}
  \includegraphics[scale=0.3]{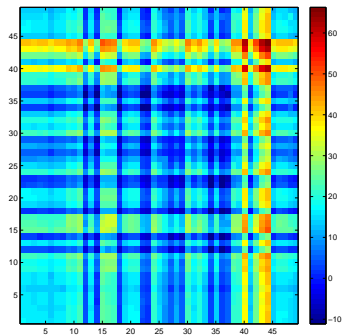}\\
  \caption{Covariance matrix which represents vector data captured from two Transfab nodes. }\label{covariance2}
\end{center}
\end{figure}

The TransFab software package is a distributed middleware developed in IBM Haifa Labs, which supports real time forwarding of message streams, providing quality of
service guarantees. We plan to use our distributed Kalman filter algorithm for online monitoring of software resources and performance. On each second each 
node records a vector of performance parameters like memory usage, CPU usage,
current bandwidth, queue sizes etc. The nodes execute the distributed Kalman filter
algorithm on the background. Figure~\ref{covariance2} plots a covariance matrix
of running an experiment using two TransFab nodes propagating data. The covariance matrix is used as an input the Kalman filter algorithm. Yellow sections show high correlation between measured parameters. Initial results are encouraging, we plan to report them using a future contribution.
\vspace{-3mm}

\section{Conclusion}
In this work we have linked together several different algorithms
from the the fields of estimation (Kalman filter),
clustering/compression (Gaussian information bottleneck) and
optimization (Affine-scaling interior-point method). Besides of
the theoretical interest in linking those different domains, we
are motivated by practical problems in communication networks. To
this end, we propose an efficient distributed iterative algorithm,
the Gaussian belief propagation algorithm, to be used for
efficiently solving these problems.

\section*{Acknowledgment}
O. Shental acknowledges the partial support of the NSF (Grant
CCF-0514859). The authors are grateful to Noam Slonim and Naftali Tishby from the
Hebrew University of Jerusalem for useful discussions.

\bibliographystyle{IEEEtran}   
\normalsize
\bibliography{IEEEabrv,final_kalman}       

\section*{Appendix A}
\begin{proof}
We prove that inverting the matrix $E$ (eq. \ref{mat_E}) is
equivalent to one iteration of the Kalman filter for computing
$P_k$.

We start from the matrix $E$ and show that $P_k^-$ can be computed
in recursion using the Schur complement formula: \BE D - CA^{-1}B
\label{Schur} \EE applied to the $2 \times 2$ upper left submatrix
of E, where $D \triangleq Q, C \triangleq A^T, B \triangleq A, A
\triangleq P_{k-1}$ we get:
\[ P_{k}^- =
\overbrace{Q}^{D} \overbrace{+}^{-}
\overbrace{A^T}^{C} \overbrace{P_{k-1}}^{-A^{-1}}
\overbrace{A}^{B}. \]

Now we compute recursively the Schur complement of lower right $2
\times 2$ submatrix of the matrix $E$ using the matrix inversion
lemma: \[ A^{-1} + A^{-1}B (D-CA^{-1}B)^{-1} CA^{-1} \]
where $A^{-1}
\triangleq P_k^-, B \triangleq H^T, C \triangleq H, D \triangleq Q
.$
In total we get:
\BE \overbrace{P_{k}^-}^{A^{-1}} +
\overbrace{P_{k}^-}^{A^{-1}}\overbrace{H^T}^{B} (\overbrace{R}^{D}
+ \overbrace{H}^{C}\overbrace{P_{k}^-}^{A^{-1}}\overbrace{H^T}^{B})^{-1}
\overbrace{H}^{C} \overbrace{P_{k}^-}^{A^{-1}} = \label{kalman_it}
\EE
\[ (I - \overbrace{P_k^-H^T(HP_k^-H^T + R)^{-1}}^{(\ref{kalman_gain})} H)P_k^- = \]
\[ = \overbrace{(I - K_kH)P_k^-}^{(\ref{p_k})} = P_k \]

\end{proof}

\section*{Appendix B}
\begin{proof}
Looking at \cite[$\S 39$]{GIB}, when $\beta = 1$ we get
\[
\begin{array}{l}
  \Sigma_{\xi+1} = (\Sigma_{t_k|y}^{-1})^{-1} =
 \Sigma_{t_k|y} = \\
 \overbrace{\Sigma_{t_k}- \Sigma_{t_ky} \Sigma_y^{-1} \Sigma_{yt_k}}^{\mbox{MMSE}} = \\
 \overbrace{\Sigma_{t_k} + B^T
\Sigma_{y|t_k} B}^{\mbox{\cite[\S 38b]{GIB}}} = \\
 \Sigma_{t_k} + \overbrace{\Sigma_{t_k}^{-1} \Sigma_{t_ky} }^{\mbox{\cite[\S
34]{GIB}}}\Sigma_{y|t_k} \overbrace{\Sigma_{yt_k}
\Sigma_{t_k}^{-1}}^{\mbox{\cite[\S 34]{GIB}}} = \\
  \overbrace{A^T \Sigma_x A + \Sigma_{\xi}}^{\mbox{\cite[\S 33]{GIB}}}
+ \overbrace{(A^T \Sigma_x A + \Sigma_{\xi})}^{\mbox{\cite[\S
33]{GIB}}} A^T \Sigma_{xy} \cdot \\
\cdot \Sigma_{y|t_k} \Sigma_{yx} A\overbrace{(A^T
\Sigma_xA + \Sigma_{\xi})^T}^{\mbox{\cite[\S 33]{GIB}}}  = \\
  A^T \Sigma_x A + \Sigma_{\xi} + (A^T \Sigma_x A + \Sigma_{\xi}) A^T \Sigma_{xy} \cdot \\
\cdot \overbrace{(\Sigma_y + \Sigma_{yt_k} \Sigma_{t_k}^{-1}
\Sigma_{t_ky})}^{\mbox{MMSE}}
\Sigma_{yx} A (A^T \Sigma_x A + \Sigma_{\xi})^T  = \\
A^T \Sigma_x A + \Sigma_{\xi}
+ (A^T \Sigma_x A + \Sigma_{\xi}) A^T \Sigma_{xy} \cdot \\
(\Sigma_y + \overbrace{A^T \Sigma_{yx}}^{\mbox{\cite[\S 5]{GIB}}}
\overbrace{(A\Sigma_xA^T + \Sigma_{\xi})}^{\mbox{(\cite[\S 5]{GIB}}}
\overbrace{\Sigma_{xy}A}^{\mbox{\cite[\S 5]{GIB}}}) \cdot \\
\cdot \Sigma_{yx} A (A^T \Sigma_x A + \Sigma_{\xi})^T. \\
\end{array}
\]

Now we show this formulation is equivalent to the Kalman filter
with the following notations: \[  P_k^- \triangleq (A^T \Sigma_x A
+ \Sigma_{\xi}) \ \ , H \triangleq A^T \Sigma_{yx},\ \   R
\triangleq \Sigma_y,\ \ \ \] \[ P_{k-1} \triangleq \Sigma_x,
 Q \triangleq \Sigma_{\xi}. \]
 Substituting we get:
\[
\begin{array}{l}
 \overbrace{(A^T \Sigma_x A + \Sigma_{\xi})}^{P_k^-} +
\overbrace{(A^T \Sigma_x A + \Sigma_{\xi})}^{P_k^-}
\overbrace{A^T \Sigma_{xy}}^{H^T} \cdot \\
  \cdot (\overbrace{\Sigma_y}^{R} + \overbrace{A^T
\Sigma_{yx}}^{H}\overbrace{(A^T \Sigma_x A +
\Sigma_{\xi})}^{P_k^-}\overbrace{\Sigma_{xy}A}^{H^T}) \cdot \\
\cdot \overbrace{\Sigma_{yx}A}^{H}\overbrace{(A^T \Sigma_x A +
\Sigma_{\xi})}^{P_k^-}. \\
\end{array}
\]
Which is equivalent to (\ref{kalman_it}). Now we can apply Theorem
1 and get the desired result.
\end{proof}

\section*{Appendix C}
\begin{proof}
In the case where $\beta > 1$, the MAP covariance matrix as
computed by the GIB algorithm is: \BE \Sigma_{\xi_{k+1}} = \beta
\Sigma_{t_k|y} + (1 - \beta) \Sigma_{t_k}
\label{weighted-GIB-kalman} \EE This is a weighted average of two
covariance matrices. $\Sigma_{t_k}$ is computed at the first phase
of the algorithm (equivalent to the prediction phase in Kalman
literature), and $\Sigma_{t_k|y}$ is computed in the second phase
of the algorithm (measurement phase). At the end of the Kalman
iteration, we simply compute the weighted average of the two
matrices to get (\ref{weighted-GIB-kalman}). Finally, we compute
$A_{k+1}$ using (eq. \ref{GIB_A}) by substituting the modified
$\Sigma_{\xi_{k+1}}$.
\end{proof}

\section*{Appendix D}
\begin{proof}
We start by expanding the Affine-scaling update rule:\\
 \[ \begin{array}{c}
 \vx_1 = \overbrace{\vx_0 -
\frac{\alpha}{\gamma}D^2\vr}^{(\ref{af_11})} = \vx_0 -
\frac{\alpha}{\underbrace{\max_i \ve_i P D
\vc}_{(\ref{af_9d})}}D^2\vr =\\
= \vx_0 - \frac{\alpha}{\max_i \ve_i \underbrace{(I -
DA^T(AD^2A^T)AD)}_{(\ref{af_10})} D \vc}D^2\vr =\\ %
= \vx_0 - \frac{\alpha D^2 \overbrace{(\vc - A^T
\vw)}^{(\ref{af_9b})}}{\max_i \ve_i (I - DA^T(AD^2A^T)^{-1}AD) D \vc} = \\
\vx_0 - \frac{\alpha D^2 (\vc - A^T
\overbrace{(AD^2A^T)^{-1}AD^2\vc}^{(\ref{af_9c})})}{\max_i \ve_i (I
- DA^T(AD^2A^T)AD)^{-1} D \vc} =\\
\vx_0 - \frac{\alpha D (I - DA^T
(AD^2A^T)^{-1}AD )D \vc}{\max_i \ve_i (I - DA^T(AD^2A^T)^{-1}AD) D
\vc}
\\
\end{array}
\]
Looking at the numerator and using the Schur complement formula
(\ref{Schur}) with the following notations: $A \triangleq
(AD^2A^T)^{-1}, B \triangleq AD, C \triangleq DA^T, D \triangleq
I$ we get the following matrix:
$ \left(%
\begin{array}{cc}
  AD^2A^T & AD \\
  DA^T & I \\
\end{array}%
\right) $. Again, the upper left block is a Schur complement $A
\triangleq 0, B \triangleq AD, C \triangleq DA^T, D \triangleq I$
of the following matrix:
$ \left(%
\begin{array}{cc}
  0 & AD \\
  DA^T & I \\
\end{array}%
\right) $. In total with get a $3 \times 3$ block matrix of the
form:
$ \left(%
\begin{array}{ccc}
  0 & AD & 0 \\
  DA^T & I & AD \\
  0 & DA^T & I \\
\end{array}%
\right)$.\\
Note that the divisor is a scalar which affects the scaling of the
step size.

Using Theorem 1, we get a computation of Kalman filter with the
following parameters: $A,H \triangleq AD, Q \triangleq I, R
\triangleq I, P_0 \triangleq 0$. This has an
interesting interpretation in the context of Kalman filter: both
prediction and measurement transformation are identical and equal
$AD$. The noise variance of both transformations are Gaussian
variables with prior $\propto \mathcal{N}(0, I)$.
\end{proof}

\end{document}